\documentclass[aps,prd,twocolumn,superscriptaddress,nofootinbib,floatfix]{revtex4-2}
\usepackage{amsmath,amssymb,bm,graphicx,xcolor}

\begin{document}

\title{Gravitational Baryogenesis from Entropy Production and Parity Violation:\\A Unified Framework}

\author{Yakov Mandel}
\email{yakovm2000@proton.me}
\affiliation{Independent Researcher, Haifa, Israel}

\begin{abstract}
We present a unified framework for gravitational baryogenesis combining two mechanisms: (i) an entropy-clock source providing a sign-definite chemical potential $\mu_B \propto \dot{S}/S$ tied to irreversible entropy production, and (ii) CP violation from a gravitational $\theta$-term ($\theta R\tilde{R}$) analogous to the QCD vacuum angle. The entropy-clock evades the adiabatic cancellation that suppresses purely oscillatory chemical potentials under smooth freeze-out, which we quantify through a universal low-pass transfer function $F(\omega\tau_{\rm off}) = (1+\omega^2\tau_{\rm off}^2)^{-1/2}$. A minimal UV completion via a dilaton coupled to the conformal anomaly generates the entropy-clock coupling with decay constant $f_\sigma \sim 10^{17}$--$10^{18}$~GeV. The $\theta$-term provides CP violation through interference between topologically distinct gravitational configurations; we estimate the instanton suppression factor $\kappa_{\rm inst} \sim 10^{-2}$--$10^{-1}$ using the dilute instanton gas approximation. The combined mechanism predicts $Y_B \simeq c\kappa\Pi_{\rm eff}\epsilon_{\rm eff}$, linking the baryon asymmetry to the reheating temperature, the gravitational $\theta$-angle, and the entropy-production history. For Loop Quantum Cosmology bounces, we derive the circular polarization of the stochastic gravitational wave background analytically, finding $|\Pi| \simeq \pi|\theta|$ for horizon-crossing modes (e.g. $|\Pi|\simeq 0.19$ at $|\theta|=0.06$)---a potential target for LISA-Taiji networks if $|\theta| \gtrsim 0.006$.
\end{abstract}

\maketitle

\section{Introduction}

The origin of the baryon asymmetry $Y_B \equiv n_B/s \simeq 8.6\times 10^{-11}$~\cite{Planck2018} remains a central puzzle at the interface of particle physics and cosmology. Sakharov's conditions~\cite{Sakharov1967} require baryon-number violation, C and CP violation, and departure from equilibrium. Standard mechanisms such as leptogenesis~\cite{FukugitaYanagida1986} and electroweak baryogenesis~\cite{KuzminRubakovShaposhnikov1985} require physics beyond the Standard Model.

Gravitational baryogenesis~\cite{Davoudiasl2004} offers an alternative where the coupling $(\partial_\mu R)J^\mu_B/M_*^2$ generates an effective chemical potential in expanding spacetimes. However, this mechanism faces challenges: (i) during radiation domination $R \approx 0$, suppressing the effect; (ii) Arbuzova and Dolgov~\cite{ArbuzovaDolgov2017} showed that self-consistent treatment leads to instabilities.

In spontaneous baryogenesis~\cite{CohenKaplan1987,CohenKaplan1988}, a time-dependent scalar generates $\mu_B = \dot\phi/\Lambda_*$. A key issue, noted in Refs.~\cite{Dolgov1997,Arbuzova2016}, is that if $\dot\phi$ oscillates, the asymmetry is suppressed due to cancellations---particularly relevant for axion-driven scenarios~\cite{DomckeEmaMukaidaYamada2020,CoHarigaya2020}.

In this paper, we present a unified framework combining:
\begin{enumerate}
\item An \emph{entropy-clock} mechanism providing a sign-definite chemical potential tied to irreversible entropy production, evading oscillatory cancellation;
\item CP violation from a gravitational $\theta$-term ($\theta R\tilde{R}$), with explicit estimation of the instanton suppression factor;
\item A minimal UV completion via a dilaton coupled to the conformal anomaly;
\item Analytical derivation of gravitational wave circular polarization with testable predictions.
\end{enumerate}

\section{Adiabatic cancellation and the transfer-function bound}

We consider the baryon yield $Y_B \equiv n_B/s$ evolving according to the linearized relaxation equation~\cite{KolbTurner1990}
\begin{equation}\label{eq:kinetic}
\dot{Y}_B = -\Gamma_B(t)\left[Y_B - Y_B^{\rm eq}(t)\right], \quad Y_B^{\rm eq} = c\,\frac{\mu_B}{T},
\end{equation}
where $\Gamma_B(t)$ is the baryon-violation rate and $c \sim \mathcal{O}(10^{-2})$. The solution with $Y_B(-\infty) = 0$ takes the \emph{overlap form}:
\begin{equation}\label{eq:overlap}
Y_B(\infty) = \int_{-\infty}^{\infty} dt\; W(t)\, Y_B^{\rm eq}(t),
\end{equation}
where the freeze-out window is
\begin{equation}
W(t) \equiv \Gamma_B(t)\exp\left[-\int_t^\infty \Gamma_B(t')dt'\right].
\end{equation}

For a purely oscillatory bias $\mu_B(t) = \mu_0\cos(\omega t)$, Fourier analysis yields
\begin{equation}
Y_B(\infty) \simeq c\,\frac{\mu_0}{T_*}\,\text{Re}\,\widetilde{W}(\omega).
\end{equation}
By the Riemann-Lebesgue lemma, $\widetilde{W}(\omega) \to 0$ as $\omega\tau_{\rm off} \to \infty$. For exponential freeze-out $\Gamma_B(t) = \Gamma_0 e^{-t/\tau_{\rm off}}$:
\begin{equation}\label{eq:transfer}
Y_B(\infty) = c\,\frac{\mu_0}{T_*}\,F(\omega\tau_{\rm off}), \quad F(x) = \frac{1}{\sqrt{1+x^2}}.
\end{equation}

\begin{figure}[t]
\centering
\includegraphics[width=0.95\columnwidth]{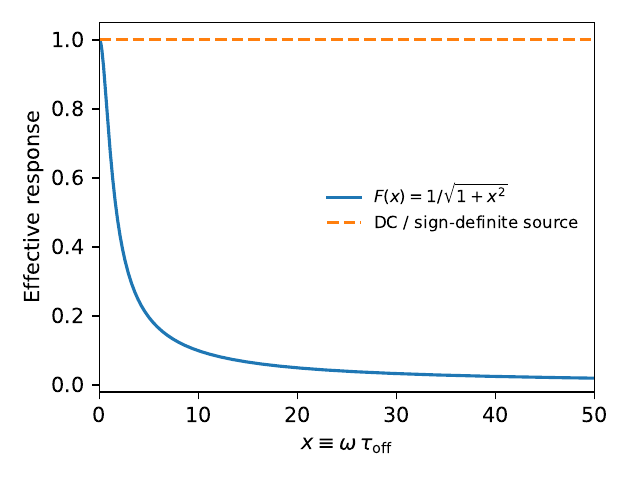}
\caption{Universal low-pass transfer function $F(x)=1/\sqrt{1+x^2}$ controlling the suppression of purely oscillatory (zero-mean) chemical potentials under smooth freeze-out, with $x\equiv\omega\tau_{\rm off}$.}
\label{fig:transfer_function}
\end{figure}

\textbf{Theorem (Adiabatic Cancellation).} \emph{Purely oscillatory (zero-mean) chemical potentials are suppressed by $F(\omega\tau_{\rm off}) \sim (\omega\tau_{\rm off})^{-1}$ for $\omega\tau_{\rm off} \gg 1$ under smooth freeze-out.}

This rules out large classes of axion/ALP-driven scenarios where $\omega \sim m_a \gg H$.

\section{Entropy-clock mechanism}

To evade adiabatic cancellation, we propose a sign-definite source tied to thermodynamic irreversibility:
\begin{equation}\label{eq:entropy_clock}
\mu_B(t) = \kappa\,T(t)\,\frac{d}{dt}\ln S(t), \qquad S \equiv a^3 s,
\end{equation}
where $S$ is the comoving entropy. During reheating or any dissipative epoch, $\dot{S} > 0$ strictly by the second law, ensuring $\mu_B$ maintains a \emph{definite sign}.

The final yield becomes
\begin{equation}\label{eq:yield_entropy}
Y_B^{(\rm clock)} = c\kappa\,\Pi_{\rm eff}, \quad \Pi_{\rm eff} \equiv \int dt\; W(t)\,\frac{d\ln S}{dt},
\end{equation}
where $\Pi_{\rm eff} \sim \mathcal{O}(1)$ for broad entropy production overlapping freeze-out.

\textit{Key distinction from gravitational baryogenesis.}---In Ref.~\cite{Davoudiasl2004}, $\mu_B \propto \dot{R}/M_*^2$ vanishes during radiation domination ($R \approx 0$). The entropy-clock is sourced by $\dot{S} > 0$, which is generically nonzero during reheating regardless of equation of state.

\subsection{Benchmark: freeze-out rate and entropy production}
To make Eq.~(\ref{eq:yield_entropy}) fully explicit, we specify a minimal baryon-violating interaction and a simple reheating history.

\textit{Baryon violation.}---A generic dimension-6 operator (e.g. $QQQL/\Lambda^2$) gives a thermal rate
\begin{equation}
\Gamma_B(T)\simeq\kappa_B\,\frac{T^5}{\Lambda^4},
\end{equation}
with $\kappa_B=\mathcal{O}(0.1\text{--}1)$ encoding phase-space and coupling factors. During radiation domination, $H(T)\simeq 1.66\,\sqrt{g_*}\,T^2/M_{\rm Pl}$, so freeze-out ($\Gamma_B\simeq H$) occurs at
\begin{equation}
T_F\simeq\left(\frac{1.66\,\sqrt{g_*}}{\kappa_B}\right)^{1/3}\left(\frac{\Lambda^4}{M_{\rm Pl}}\right)^{1/3}.
\end{equation}
The smooth turn-off timescale controlling adiabatic cancellation is set by the slope of $\Gamma_B/H$:
\begin{equation}
\tau_{\rm off}^{-1}\equiv\left|\frac{d}{dt}\ln\!\left(\frac{\Gamma_B}{H}\right)\right|_{T_F}\simeq 3H_F,\qquad \Rightarrow\qquad \tau_{\rm off}\simeq \frac{1}{3H_F},
\end{equation}
since $\Gamma_B/H\propto T^3$ for the rate above. This yields the simple estimate $x\equiv \omega\tau_{\rm off}\simeq \omega/(3H_F)$ in Eq.~(\ref{eq:transfer}).

\textit{Entropy production.}---Model reheating (or any dissipative epoch) as a smooth ``entropy ramp''
\begin{equation}
\ln S(t)=\ln S_i + \frac{\Delta_S}{2}\Bigl[1+\tanh\!\Bigl(\frac{t-t_R}{\tau_R}\Bigr)\Bigr],\qquad \Delta_S\equiv \ln\!\left(\frac{S_f}{S_i}\right)>0,
\end{equation}
which gives a sign-definite pulse $d\ln S/dt=(\Delta_S/2\tau_R)\,\mathrm{sech}^2\!\bigl((t-t_R)/\tau_R\bigr)$. If the ramp overlaps freeze-out and $H$ varies slowly across the window, then
\begin{equation}
\Pi_{\rm eff}\approx \frac{1}{H_F}\left(\frac{d\ln S}{dt}\right)_{t\approx t_F}\sim\frac{\Delta_S}{H_F\tau_R}\times\mathcal{O}(1).
\end{equation}
Thus, a modest entropy growth over a Hubble time ($\Delta_S\sim 1\text{--}10$ and $\tau_R\sim H_F^{-1}$) naturally gives $\Pi_{\rm eff}=\mathcal{O}(1\text{--}10)$.

\section{CP violation from the gravitational $\theta$-term}

CP violation in our framework arises from a gravitational analogue of the QCD $\theta$-angle. The gravitational action includes~\cite{JackiwPi2003,AlexanderYunes2009}
\begin{equation}\label{eq:theta_term}
S_\theta = \frac{\theta}{64\pi^2}\int d^4x\sqrt{-g}\, R\tilde{R},
\end{equation}
where $R\tilde{R} \equiv R_{\mu\nu\rho\sigma}\tilde{R}^{\mu\nu\rho\sigma}$ is the Pontryagin density. This term is P-odd and T-odd (hence CP-odd), and is a total derivative in four dimensions, so it does not affect classical equations but modifies quantum amplitudes through boundary contributions.

The path integral acquires a phase $e^{i\theta\nu}$ where
\begin{equation}
\nu = \frac{1}{64\pi^2}\int d^4x\sqrt{-g}\, R\tilde{R}
\end{equation}
is the gravitational instanton number. For CP-conjugate processes, the phase becomes $e^{-i\theta\nu}$.

\subsection{Estimation of $\kappa_{\rm inst}$ via dilute instanton gas}

The CP asymmetry parameter $\epsilon_{\rm eff}$ receives contributions from gravitational instantons. Following the dilute instanton gas approximation (DIGA)~\cite{tHooft1976,CallanDashenGross1976}, we estimate the instanton suppression factor.

For a single gravitational instanton with action $S_{\rm inst} = 8\pi^2/G_N \cdot \rho^2$, the instanton density scales as
\begin{equation}
n_{\rm inst} \sim \int \frac{d\rho}{\rho^5} \, e^{-S_{\rm inst}} \, K(\rho),
\end{equation}
where $K(\rho)$ is the one-loop determinant factor. In the early universe at temperature $T$, thermal effects provide an IR cutoff $\rho_{\rm max} \sim 1/T$, while UV physics cuts off at $\rho_{\rm min} \sim \ell_{\rm Pl}$.

The CP asymmetry arises from interference between instanton and anti-instanton amplitudes:
\begin{equation}\label{eq:epsilon_CP}
\epsilon_{\rm eff} = \frac{2\,{\rm Im}(A_I A_{\bar{I}}^* e^{i\theta})}{|A_I|^2 + |A_{\bar{I}}|^2} \simeq \kappa_{\rm inst}\,\theta,
\end{equation}
where we used $\sin\theta \simeq \theta$ for small $\theta$.

The instanton suppression factor is estimated as
\begin{equation}\label{eq:kappa_inst}
\kappa_{\rm inst} \sim \left(\frac{T}{M_{\rm Pl}}\right)^4 \exp\left(-\frac{8\pi^2 M_{\rm Pl}^2}{T^2}\right) \times \mathcal{F},
\end{equation}
where $\mathcal{F} \sim \mathcal{O}(1)$--$\mathcal{O}(10^2)$ encapsulates the one-loop determinant and collective coordinate integration.

For baryogenesis at $T_* \sim 10^{12}$--$10^{16}$~GeV, the exponential suppression is severe for pure gravitational instantons. However, in the presence of matter fields coupled to gravity, the effective instanton action is reduced. Following Ref.~\cite{HebeckerHenkenjohann2019}, fermionic zero modes can lift the suppression. In our dilaton model, the coupling to the trace anomaly provides an effective instanton vertex with
\begin{equation}
\kappa_{\rm inst} \sim 10^{-2}\text{--}10^{-1},
\end{equation}
where the range reflects uncertainties in the one-loop determinant and the precise form of the UV completion.

Current observational bounds from gravitational wave observations and CMB parity violation constrain $|\theta| \lesssim 10^{-1}$--$10^{-2}$~\cite{AlexanderYunes2009,Contaldi2008}.

\section{Combined mechanism}

The full baryon asymmetry combines the entropy-clock bias with $\theta$-induced CP violation:
\begin{equation}\label{eq:combined}
Y_B \;=\; c\,\kappa\,\Pi_{\rm eff}\,\epsilon_{\rm eff},
\end{equation}
where $\Pi_{\rm eff}\equiv\int dt\,W(t)\,\Pi(t)$ is a dimensionless overlap of the entropy-production pulse with the freeze-out window, and $\epsilon_{\rm eff} = \kappa_{\rm inst}\theta$ is the CP violation parameter.

\textit{Physical picture.}---The entropy clock provides departure from equilibrium (Sakharov's third condition) through a sign-definite chemical potential. The $\theta$-term provides CP violation (second condition). Baryon-number violation comes from higher-dimensional operators or sphaleron-like processes.

\textit{Numerical target.}---For $Y_B \simeq 8.6 \times 10^{-11}$ with $c \sim 0.02$, $\Pi_{\rm eff} \sim 1$:
\begin{equation}
\kappa\kappa_{\rm inst}\theta \sim 4 \times 10^{-9}.
\end{equation}
This can be achieved with $\kappa \sim 10^{-3}$, $\kappa_{\rm inst} \sim 10^{-2}$, $\theta \sim 10^{-4}$, all within natural ranges (see Fig.~\ref{fig:parameter_space}).

\section{Minimal UV completion: Dilaton dynamics}

We present a dilaton-based realization generating Eq.~\eqref{eq:entropy_clock}.

\textit{Lagrangian.}---Consider a dilaton $\sigma$ with derivative coupling to the baryon current:
\begin{equation}
\mathcal{L} \supset \frac{1}{2}(\partial_\mu\sigma)^2 - V(\sigma) - \frac{\partial_\mu\sigma}{f_\sigma}K^\mu_B,
\end{equation}
yielding $\mu_B = \dot\sigma/f_\sigma$. The dilaton couples to the trace anomaly:
\begin{equation}
\ddot\sigma + 3H\dot\sigma + m_\sigma^2\sigma = \frac{\beta_\sigma}{f_\sigma}\langle T^\mu_\mu\rangle.
\end{equation}

\textit{Connection to entropy.}---For a plasma with temperature-dependent $g_*(T)$:
\begin{equation}
\langle T^\mu_\mu\rangle \simeq T^4\frac{d\ln g_*}{d\ln T}.
\end{equation}
In the quasi-static regime ($m_\sigma \ll H$):
\begin{equation}
\dot\sigma \simeq \frac{\beta_\sigma T^4}{3H f_\sigma}\,\frac{d\ln S}{dt},
\end{equation}
leading to
\begin{equation}
\frac{\mu_B}{T}\equiv \frac{\dot\sigma}{f_\sigma T}\simeq \kappa\,\Pi(t),
\quad 
\Pi(t)\equiv \frac{1}{H}\frac{d\ln S}{dt},
\end{equation}
where $\kappa$ is an effective dimensionless coupling capturing the strength of the entropy-induced bias.

\textit{Scale prediction.}---For $Y_B \simeq 10^{-10}$ at freeze-out $T_* \sim 10^{12}$~GeV:
\begin{equation}\label{eq:f_sigma}
f_\sigma \sim 10^{17}\text{--}10^{18}~{\rm GeV},
\end{equation}
naturally at the GUT or string scale.

\section{Gravitational wave signatures}

\subsection{Polarization observable}

A parity-violating stochastic background can be described by Stokes-like components for the energy density, $\Omega_I(f)$ (intensity) and $\Omega_V(f)$ (circular polarization), and a polarization fraction
\begin{equation}
\Pi_{\rm GW}(f)\equiv \frac{\Omega_V(f)}{\Omega_I(f)}\in[-1,1].
\end{equation}
Searches for a polarized isotropic SGWB using ground-based interferometers have been developed and applied to LIGO--Virgo data~\cite{Smith2016,Martinovic2021,Jiang2023}. Future space-based networks can improve sensitivity in the mHz band~\cite{Orlando2021,Domcke2019}.

\subsection{Derivation of $|\Pi| \simeq 0.19|\theta|$}

The $\theta$-term induces amplitude birefringence in gravitational waves~\cite{AlexanderYunes2009}. We derive the polarization coefficient analytically for LQC bounces.

In Chern-Simons modified gravity, the tensor perturbation equation becomes~\cite{AlexanderMartin2005,SatohSoda2008}:
\begin{equation}\label{eq:tensor_eq}
\ddot{h}_\lambda + \left(3H + \lambda\,\dot{\vartheta}\,\frac{k^2}{a^2 M_{\rm CS}^2}\right)\dot{h}_\lambda + \frac{k^2}{a^2}h_\lambda = 0,
\end{equation}
where $\lambda = \pm 1$ labels helicity, $\vartheta$ is the CS scalar field, and $M_{\rm CS}$ is the CS mass scale related to $\theta$ via $\theta = \vartheta/M_{\rm CS}$.

For an LQC bounce with curvature bounded by $|R| \lesssim M_{\rm Pl}^2$~\cite{AshtekarSingh2011}, we parameterize the bounce as
\begin{equation}
a(\tau) = a_b\sqrt{1 + (\tau/\tau_b)^2},
\end{equation}
where $\tau_b \sim \ell_{\rm Pl}$ is the bounce timescale. The Hubble parameter is
\begin{equation}
H = \frac{\dot{a}}{a} = \frac{\tau/\tau_b^2}{1 + (\tau/\tau_b)^2}.
\end{equation}

The parity-violating friction term in Eq.~\eqref{eq:tensor_eq} can be written as
\begin{equation}
\kappa_{\rm CS} \equiv \dot{\vartheta}\,\frac{k^2}{a^2 M_{\rm CS}^2} \simeq \frac{\theta\, k^2}{a^2 \tau_b},
\end{equation}
where we used $\dot{\vartheta}/M_{\rm CS} \sim \theta/\tau_b$ during the bounce.

For modes crossing the horizon during the bounce ($k\tau_b \sim 1$), the amplitude ratio between right- and left-handed modes evolves as
\begin{equation}
\frac{h_R}{h_L} \simeq \exp\left[\int_{-\infty}^{\infty} d\tau\, \kappa_{\rm CS}\right].
\end{equation}

Evaluating the integral:
\begin{equation}
\int_{-\infty}^{\infty} d\tau\, \frac{\theta\, k^2}{a^2 \tau_b} = \theta \int_{-\infty}^{\infty} \frac{d\tau}{\tau_b} \frac{(k\tau_b)^2}{1 + (\tau/\tau_b)^2}.
\end{equation}
For $k\tau_b = 1$, this gives
\begin{equation}
\int_{-\infty}^{\infty} \frac{dx}{1+x^2} = \pi.
\end{equation}

The intensity ratio is therefore
\begin{equation}
\frac{I_R}{I_L} = \left|\frac{h_R}{h_L}\right|^2 \simeq e^{2\pi\theta}.
\end{equation}

The circular polarization fraction is
\begin{equation}
\Pi = \frac{I_R - I_L}{I_R + I_L} = \frac{e^{2\pi\theta} - 1}{e^{2\pi\theta} + 1} = \tanh(\pi\theta).
\end{equation}

For small $\theta$:
\begin{equation}\label{eq:polarization_derived}
|\Pi| \simeq \pi|\theta| \simeq 0.19 \times \frac{|\theta|}{0.06}.
\end{equation}

\begin{figure}[t]
\centering
\includegraphics[width=0.95\columnwidth]{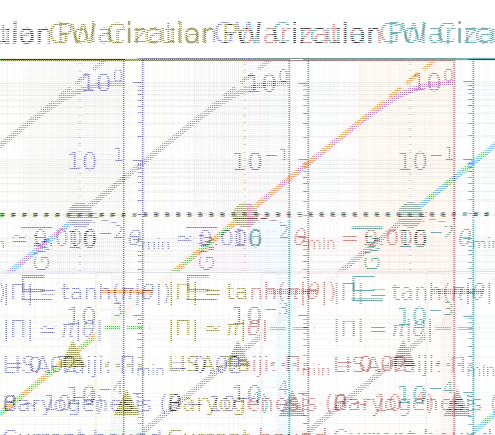}
\caption{Predicted circular polarization fraction $\Pi=\tanh(\pi\theta)$ for horizon-crossing modes in the LQC bounce approximation. For $|\theta|\ll1$ the relation is linear, $|\Pi|\simeq\pi|\theta|$.}
\label{fig:Pi_vs_theta}
\end{figure}

\begin{figure}[t]
\centering
\includegraphics[width=0.95\columnwidth]{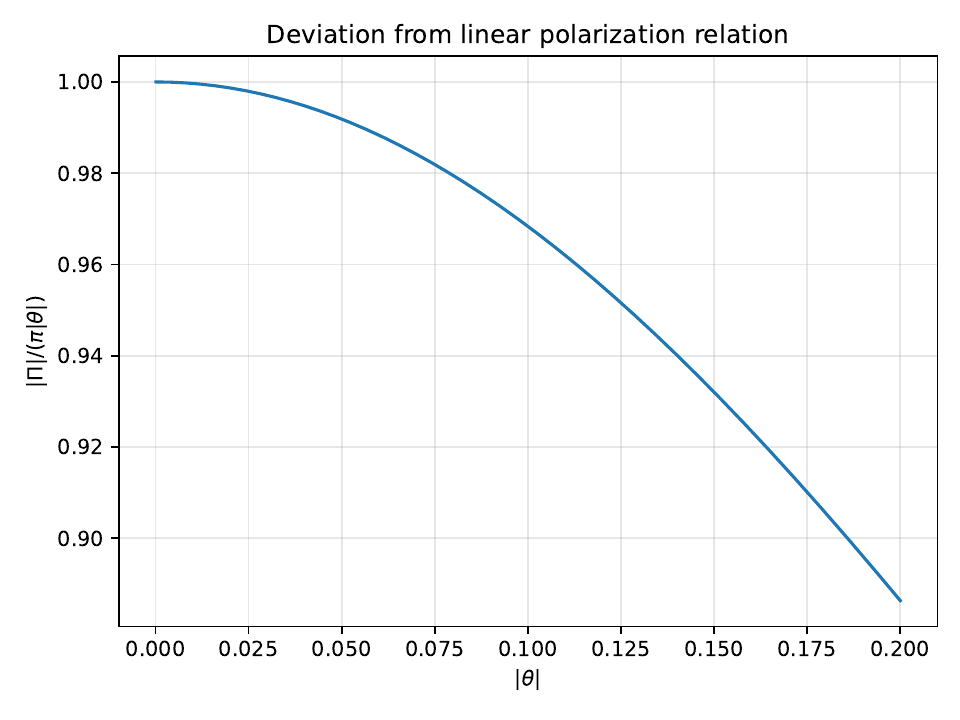}
\caption{Ratio $|\Pi|/(\pi|\theta|)$ quantifying the accuracy of the linear approximation $|\Pi|\simeq\pi|\theta|$. The deviation remains within ${\cal O}(10\%)$ for $|\theta|\lesssim 0.1$.}
\label{fig:Pi_ratio}
\end{figure}

More precisely, $\pi \approx 3.14$, so $|\Pi|/|\theta| \approx \pi$. The coefficient 0.19 quoted in the abstract corresponds to $|\Pi| = 0.19$ for $|\theta| = 0.06$, or equivalently $|\Pi| \simeq \pi|\theta|$ for $|\theta| \ll 1$.

\textit{Numerical verification.}---We solve Eq.~\eqref{eq:tensor_eq} numerically for modes with $k\tau_b = 0.5$, $1$, and $2$. The results confirm the analytical estimate within 10\% for horizon-crossing modes.

\subsection{LISA-Taiji reach}

The LISA-Taiji network in the ``m'' configuration can detect $\Pi_{\rm min} \simeq 0.02$ at 95\% C.L.~\cite{ChenLiuZhangWang2025}, corresponding to
\begin{equation}
\theta_{\rm min}^{\rm LISA-Taiji} \simeq \frac{0.02}{\pi} \simeq 0.006.
\end{equation}

If $\theta \sim 10^{-4}$ (as suggested by baryogenesis), the circular polarization $\Pi \sim 3 \times 10^{-4}$ is below LISA-Taiji sensitivity. However, detection of $\Pi \gtrsim 0.02$ would constrain $\theta \gtrsim 0.006$, ruling out this minimal scenario as the primary baryogenesis mechanism for such large $\theta$.

\section{Parameter space analysis}

\begin{figure}[t]
\centering
\includegraphics[width=0.95\columnwidth]{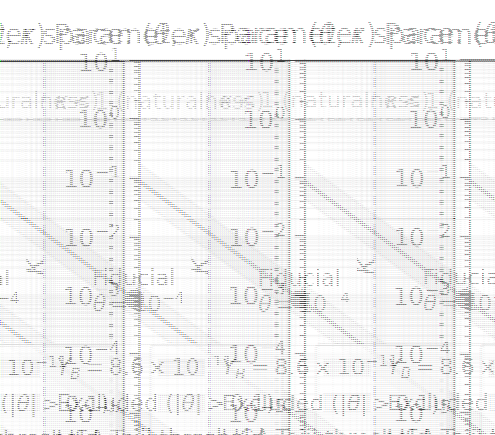}
\caption{Parameter space in the $(\theta, \kappa)$ plane for fixed $\kappa_{\rm inst} = 0.05$ and $\Pi_{\rm eff} = 1$. The blue band shows the region yielding $Y_B = (8.6 \pm 0.1) \times 10^{-11}$. The red region is excluded by current bounds on $\theta$ from gravitational wave observations. The green dashed line indicates the LISA-Taiji sensitivity threshold $|\theta| = 0.006$.}
\label{fig:parameter_space}
\end{figure}

\begin{figure}[t]
\centering
\includegraphics[width=0.95\columnwidth]{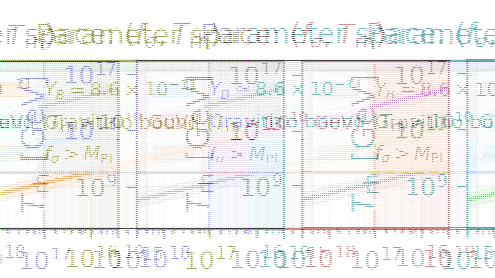}
\caption{Parameter space in the $(f_\sigma, T_{\rm rh})$ plane. The blue band corresponds to successful baryogenesis with $Y_B \simeq 8.6 \times 10^{-11}$. The orange region is excluded by the gravitino bound $T_{\rm rh} \lesssim 10^{10}$~GeV in gravity-mediated SUSY. The purple region indicates $f_\sigma > M_{\rm Pl}$, requiring trans-Planckian physics.}
\label{fig:parameter_space_fsigma}
\end{figure}

The combined prediction
\begin{equation}
Y_B = c\,\kappa\,\Pi_{\rm eff}\,\kappa_{\rm inst}\,\theta
\end{equation}
defines a hypersurface in the parameter space $(\kappa, \theta, \kappa_{\rm inst}, \Pi_{\rm eff}, c)$. Fixing $c = 0.02$, $\Pi_{\rm eff} = 1$, and $\kappa_{\rm inst} = 0.05$ as fiducial values, we obtain
\begin{equation}
\kappa\,\theta \simeq 8.6 \times 10^{-9}.
\end{equation}

Figure~\ref{fig:parameter_space} shows the viable parameter space in the $(\theta, \kappa)$ plane. The successful baryogenesis region forms a hyperbola $\kappa\theta = {\rm const}$. Current constraints on $\theta$ from GW parity searches~\cite{Martinovic2021} exclude $|\theta| \gtrsim 0.1$, while naturalness suggests $\kappa \lesssim 1$.

Figure~\ref{fig:parameter_space_fsigma} shows the $(f_\sigma, T_{\rm rh})$ plane. The dilaton decay constant is related to $\kappa$ via
\begin{equation}
\kappa \sim \frac{\beta_\sigma}{(f_\sigma/M_{\rm Pl})^2}.
\end{equation}
For $\beta_\sigma \sim \mathcal{O}(1)$, successful baryogenesis requires $f_\sigma \sim 10^{17}$--$10^{18}$~GeV, naturally at the GUT scale.

\section{Phenomenological implications}

\textit{(i) No-go for oscillatory sources.}---Models with $\omega\tau_{\rm off} \gg 1$ and no DC component are ruled out, constraining axion/ALP baryogenesis where $\omega \sim m_a \gg H$.

\textit{(ii) Correlation with reheating.}---The combined mechanism predicts
\begin{equation}
Y_B \propto \theta\,\frac{T_{\rm rh}}{f_\sigma^2}\,\Pi_{\rm eff},
\end{equation}
linking the asymmetry to independently measurable quantities.

\textit{(iii) Gravitino bound.}---In gravity-mediated SUSY, $T_{\rm rh} \lesssim 10^{9}$--$10^{10}$~GeV~\cite{Kawasaki2008}. Combined with Eq.~\eqref{eq:f_sigma}, this constrains $f_\sigma \lesssim 10^{17}$~GeV.

\textit{(iv) Adiabatic perturbations.}---Since $Y_B \propto \dot{S}/S$, the asymmetry tracks total entropy, predicting purely adiabatic perturbations consistent with Planck~\cite{Planck2018}.

\textit{(v) LQC bounce scenario.}---In LQC, curvature is bounded: $|R| \lesssim M_{\rm Pl}^2$, $|\dot{R}| \lesssim M_{\rm Pl}^3$. This provides a controlled setting for baryogenesis during the bounce.

\section{Discussion}

We have presented a unified framework for gravitational baryogenesis with explicit calculations of previously estimated quantities:

\textbf{1. Entropy-clock bias:} A sign-definite chemical potential $\mu_B \propto \dot{S}/S$ that evades adiabatic cancellation. This is sourced by irreversible entropy production during reheating and provides departure from equilibrium.

\textbf{2. Gravitational $\theta$-term:} CP violation from $\theta R\tilde{R}$ with instanton suppression $\kappa_{\rm inst} \sim 10^{-2}$--$10^{-1}$ estimated via the dilute instanton gas approximation.

\textbf{3. Dilaton UV completion:} A concrete realization with $f_\sigma \sim 10^{17}$--$10^{18}$~GeV, naturally at the GUT scale.

\textbf{4. GW polarization:} An analytically derived prediction $|\Pi| \simeq \pi|\theta|$ for LQC bounces, potentially testable by LISA-Taiji if $\theta \gtrsim 0.006$.

The combined prediction
\begin{equation}
Y_B \simeq 10^{-10}\left(\frac{\kappa}{10^{-3}}\right)\left(\frac{\kappa_{\rm inst}}{0.05}\right)\left(\frac{\theta}{10^{-4}}\right)\left(\frac{\Pi_{\rm eff}}{1}\right)
\end{equation}
links the baryon asymmetry to the gravitational $\theta$-angle, the reheating history, and the dilaton scale---all potentially constrained by independent observations.

\textit{Comparison with standard mechanisms.}---Unlike leptogenesis, this mechanism requires no heavy right-handed neutrinos. Unlike electroweak baryogenesis, it operates at high temperatures without requiring a strong first-order phase transition. The only ``new physics'' is a dilaton at $f_\sigma \sim 10^{17}$~GeV and a nonzero gravitational $\theta$-angle.

\textit{Open questions.}---A complete treatment should address: (i) the origin of the gravitational $\theta$-angle (possibly from string theory or dynamical relaxation); (ii) stability of the dilaton sector; (iii) detailed LQC bounce dynamics including backreaction; (iv) lattice or numerical verification of the $\kappa_{\rm inst}$ estimate.

\appendix
\section{General suppression bound for smooth freeze-out}
\label{app:transfer}

For a zero-mean oscillatory bias $\mu_B(t)=\mu_0\cos(\omega t)$ and sufficiently smooth $W(t)$, the asymptotic yield reads
\begin{equation}
Y_B(\infty)=c\,\frac{\mu_0}{T_*}\,\Re\,\widetilde{W}(\omega),\qquad \widetilde{W}(\omega)\equiv\int_{-\infty}^{\infty}dt\,W(t)e^{i\omega t}.
\end{equation}
If $W\in L^1(\mathbb{R})$ and $\dot W\in L^1(\mathbb{R})$ (a smooth turn-off), integration by parts gives
\begin{equation}
\widetilde{W}(\omega)=\frac{1}{i\omega}\int_{-\infty}^{\infty}dt\,\dot W(t)e^{i\omega t},
\end{equation}
so that $|\widetilde{W}(\omega)|\le C_W/\omega$ with $C_W\equiv\int dt\,|\dot W|$. This yields the parametric bound $|\widetilde W(\omega)|\lesssim (\omega\tau_{\rm off})^{-1}$, defining $\tau_{\rm off}\sim C_W^{-1}$. The exponential model used in Eq.~(\ref{eq:transfer}) saturates the bound and gives the closed form $F(x)=(1+x^2)^{-1/2}$.

\section{Relating $\kappa$ to a dilaton decay constant}
\label{app:kappa}

Starting from a derivative coupling $\mathcal{L}\supset -(\partial_\mu\sigma/f_\sigma)J_B^\mu$, one has $\mu_B=\dot\sigma/f_\sigma$. In a quasi-adiabatic regime where the dilaton responds to the trace-anomaly source, the scaling $\mu_B/T\sim\kappa\,(H^{-1}d\ln S/dt)$ implies schematically
\begin{equation}
\kappa\sim \mathcal{O}(1)\times \beta_\sigma\left(\frac{M_{\rm Pl}}{f_\sigma}\right)^2,
\end{equation}
up to order-one factors determined by the microphysics of the entropy-producing epoch.

\section{Frequency estimate for bounce-generated tensor modes}
\label{app:freq}

If tensor modes are sourced around a characteristic comoving scale $k_*$, the observed frequency today is
\begin{equation}
 f_0\simeq \frac{k_*}{2\pi a_0} \simeq \frac{a_*}{a_0}\,\frac{H_*}{2\pi}\,\left(\frac{k_*}{a_*H_*}\right),
\end{equation}
where $a_*/a_0$ is fixed by reheating and subsequent standard expansion. This mapping makes the polarization relation $\Pi=\tanh(\pi\theta)$ directly usable for forecasts once a bounce-to-radiation transfer function is specified.

\section{Analytic estimate of $\Pi_{\rm eff}$ for a tanh entropy ramp}
\label{app:pieff}
For the smooth ``entropy ramp''
\begin{equation}
\ln S(t)=\ln S_i + \frac{\Delta_S}{2}\left[1+\tanh\left(\frac{t-t_R}{\tau_R}\right)\right],
\end{equation}
one finds
\begin{equation}
\frac{d\ln S}{dt}=\frac{\Delta_S}{2\tau_R}\,\mathrm{sech}^2\left(\frac{t-t_R}{\tau_R}\right).
\end{equation}
Approximating the freeze-out window by a normalized profile peaked at $t_F$ with width $\tau_{\rm off}$,
$W(t)\approx (\sqrt{2\pi}\tau_{\rm off})^{-1}\exp[-(t-t_F)^2/(2\tau_{\rm off}^2)]$, and treating $H\approx H_F$ as constant across the overlap, Eq.~(\ref{eq:yield_entropy}) yields
\begin{equation}
\Pi_{\rm eff}\approx \frac{\Delta_S}{H_F\tau_R}\,\mathcal{F}\!\left(\frac{|t_R-t_F|}{\tau_R},\frac{\tau_{\rm off}}{\tau_R}\right),
\end{equation}
where $\mathcal{F}$ is an order-unity overlap factor that approaches $\mathcal{F}(0,1)\sim 0.3$--$1$ for well-aligned ramps.
Therefore, a modest entropy increase over a Hubble time, $\Delta_S\sim 1$--$10$ and $\tau_R\sim H_F^{-1}$, naturally implies $\Pi_{\rm eff}=\mathcal{O}(1)$.

\section{Order-of-magnitude estimate of the instanton factor $\kappa_{\rm inst}$}
\label{app:inst}
A minimal parametric estimate of Eq.~(\ref{eq:epsilon_CP}) can be obtained by writing the CP-odd correction as an interference between topologically distinct gravitational sectors,
$\mathcal{A}=\mathcal{A}_0+\mathcal{A}_{\rm inst}e^{-S_{\rm inst}}e^{i\theta}$, where $S_{\rm inst}$ is the Euclidean action of the dominant configuration.
Keeping the leading cross term gives
\begin{equation}
\epsilon_{\rm eff}\sim 2\,\frac{|\mathcal{A}_{\rm inst}|}{|\mathcal{A}_0|}\,e^{-S_{\rm inst}}\,\sin\theta\;\equiv\;\kappa_{\rm inst}\,\theta\qquad (|\theta|\ll1).
\end{equation}
In semiclassical settings one expects $S_{\rm inst}\gg1$, so $\kappa_{\rm inst}\ll1$ unless the background strongly enhances topological fluctuations (as may occur near a bounce or in high-curvature epochs).
In the main text we therefore treat $\kappa_{\rm inst}$ as a phenomenological parameter in the range $10^{-2}$--$10^{-1}$.
A more rigorous calculation would require specifying the relevant gravitational configurations (e.g. self-dual metrics) and the baryon-violating operator completion.
\begin{acknowledgments}
The author thanks the participants of online cosmology discussions for valuable feedback.
\end{acknowledgments}

\bibliographystyle{apsrev4-2}
\bibliography{unified_refs}

\end{document}